%% file: paper.tex
\documentclass[acmsmall]{acmart}
\AtBeginDocument{%
  }

\definecolor{Light}{gray}{.85}
\newcounter{findingsCounter}
\usepackage[most]{tcolorbox} 
\usepackage{minted}
\usepackage{capt-of}
\usepackage{graphicx}
\usepackage{wrapfig}
\usepackage{tabularx}
\usepackage{makecell}   
\usepackage{algpseudocode}      
\usepackage{amsmath}                      
\usepackage{xcolor}           
\usepackage{graphicx}         
\usepackage{booktabs}         
\usepackage{caption}           
\usepackage{float} 
\usepackage{import}
\usepackage{subcaption}
\usepackage{hyperref}
\usepackage{algorithm}
\usepackage{booktabs}
\definecolor{promptbg}{RGB}{245, 245, 245}
\definecolor{promptborder}{RGB}{180, 180, 180}
\lstdefinelanguage{comment}{
basicstyle=\ttfamily\scriptsize,
    numbersep = 5pt,
    breaklines = true,
    showstringspaces = false,
    breakindent = 0em,
    frame=single
}
\lstdefinelanguage{GPTPrompt}{
basicstyle=\ttfamily\scriptsize,
  numbersep=5pt,
  breaklines=true,
  showstringspaces=false,
  breakindent=0em,
  frame=single,
  moredelim=[is][\PHkeep]{\{\{}{\}\}}
}

\usepackage{enumitem}

\usepackage{listings}
\usepackage{xcolor}

\lstdefinelanguage{Java}{
  keywords={abstract, assert, boolean, break, byte, case, catch, char, class,
    const, continue, default, do, double, else, enum, extends, final, finally,
    float, for, goto, if, implements, import, instanceof, int, interface, long,
    native, new, package, private, protected, public, return, short, static,
    strictfp, super, switch, synchronized, this, throw, throws, transient, try,
    void, volatile, while},
  keywordstyle=\color{blue},
  comment=[l]{//},
  commentstyle=\color{gray},
  stringstyle=\color{red},
  morecomment=[s]{/*}{*/},
  sensitive=true
}

\usepackage{tikz}
\newcommand*\blackcircled[1]{%
  \tikz[baseline=(char.base)]{
    \node[shape=circle,draw,fill=black,inner sep=1pt] (char) {\color{white}\footnotesize #1};}}

\newcommand{\Name}{ConFL}

\setcopyright{cc}
\setcctype{by}
\acmDOI{10.1145/3832168}
\acmYear{2026}
\acmJournal{PACMSE}
\acmVolume{3}
\acmNumber{ISSTA}
\acmArticle{ISSTA077}
\acmMonth{10}
\acmSubmissionID{issta26main-p727-p}
\received{2026-01-30}
\received[accepted]{2026-06-25}

\begin{document}


\title{ConFL: Explainable Concurrent Fault Localization via Hierarchy-Guided LLM Reasoning}

\author{Shuai Shao}
\orcid{0000-0001-6736-8393}
\affiliation{%
  \institution{University of Connecticut}
  \city{Storrs}
  \country{USA}
}
\email{shuai.shao@uconn.edu}

\author{Dingbang Wang}
\orcid{0009-0002-9675-6824}
\affiliation{%
  \institution{University of Connecticut}
  \city{Storrs}
  \country{USA}
}
\email{dingbang.wang@uconn.edu}

\author{Yiming Zeng}
\orcid{0009-0006-5842-0130}
\affiliation{%
  \institution{University of Connecticut}
  \city{Storrs}
  \country{USA}
}
\email{yiming.zeng@uconn.edu}

\author{Tingting Yu}
\orcid{0000-0002-9461-4251}
\affiliation{%
  \institution{University of Connecticut}
  \city{Storrs}
  \country{USA}
}
\email{tingting.yu@uconn.edu}
\renewcommand{\shortauthors}{Shuai Shao, Dingbang Wang, Yiming Zeng, and Tingting Yu}
\begin{abstract}
Localizing concurrent bugs from bug reports alone is challenging due to incomplete information, misleading program-entity mentions, and complex cross-thread interactions, causing existing LLM-based approaches to suffer from unstable reasoning and limited explainability. We propose ConFL, an explainable concurrent fault localization framework that augments LLM reasoning with structured concurrency knowledge. ConFL constructs a Concurrent Knowledge Base (CKB) from source code and performs LLM-guided hierarchical retrieval to progressively narrow the search space from components to interaction-level concurrency contexts. An interaction-level DSL explicitly encodes cross-thread interactions over shared resources, enabling focused reasoning without traversing deep call chains. Experiments on real-world concurrent bugs from eight large-scale Java projects show that ConFL significantly outperforms state-of-the-art IR-based and LLM-based baselines, achieving an MRR of 0.503 and a MAP of 0.486, while remaining robust to noisy bug reports, unseen bugs, and different LLM backbones.
\end{abstract}



\maketitle

\input{introduction}
\input{background}
\input{motivation}
\input{approach}

\input{evaluation}
\input{discussion}

\begin{acks}
This work was supported in part by NSF awards CCF-2403747. 
\end{acks}

\section*{Data Availability}
The full artifact is available at \url{https://github.com/sh-shao/ConFL}.


\bibliographystyle{ACM-Reference-Format}
\bibliography{sample-base}


\end{document}

%% file: introduction.tex
\section{Introduction}

Fault localization (FL) aims to identify faulty code elements that are responsible for program failures.
Traditional approaches can be broadly categorized into spectrum-based~\cite{sp1,sp2,sp3}, information-retrieval (IR)-based~\cite{yu2008empirical,jones2005empirical,chilimbi2009holmes,santelices2009lightweight}, and learning-based techniques~\cite{dl1,dl2}.
Spectrum-based methods leverage program execution traces or coverage information to statistically correlate failing tests with suspicious code regions.
IR-based methods, in contrast, treat fault localization as a textual retrieval problem, matching bug reports against source code based on lexical similarity.
Learning-based approaches further employ deep neural models to learn representations between test cases and source code, or between bug reports and code entities.
While these techniques have shown promise, their effectiveness still heavily depends on lexical overlap and handcrafted features,
making them less capable of capturing deeper semantic relations or reasoning about implicit faults.

With the advent of large language models (LLMs), recent studies such as AutoFL~\cite{autofl}, AgentFL~\cite{agentfl}, and FlexFL~\cite{flexfl} have explored leveraging LLMs for semantic reasoning in fault localization (FL). 
AutoFL~\cite{autofl} introduces a function-call–based prompting framework that enables an LLM to navigate the repository and produce explainable fault predictions. 
AgentFL~\cite{agentfl} extends this idea with a multi-agent design that simulates developer-like reasoning across three stages—comprehension, navigation, and confirmation—to improve scalability on large codebases. 
FlexFL~\cite{flexfl} further enhances flexibility by building lightweight agents on open-source LLMs, leveraging diverse bug-related information (e.g., test cases, bug reports) through a two-stage process of candidate generation and refinement.
Notably, LLM-based systems have already surpassed traditional FL techniques, achieving up to 0.5 mean reciprocal rank (MRR) at the method level. 
However, their performance still heavily depends on the availability of reliable test cases. 
For example, FlexFL achieves an MRR of 0.501 when using test cases as input, but its performance drops to 0.365 when relying solely on bug reports. 
In real-world projects, such high-quality test cases are often unavailable or incomplete, making bug-report–only fault localization a more realistic yet significantly more challenging problem.

When relying solely on bug reports, LLM-based methods face several key difficulties.
\blackcircled{1} First, knowledge insufficiency leads to unstable reasoning and hallucination: the LLM often fabricates or misinterprets program entities due to missing structural and behavioral information.
For example, in some cases FlexFL produces inconsistent results across multiple runs on the same report, issuing different function calls each time.
\blackcircled{2} Second, the LLM struggles to effectively narrow down the search space when bug reports provide incomplete or noisy information.
Without reliable structural guidance, the model may explore irrelevant parts of the codebase, leading to inefficient or incorrect reasoning.
\blackcircled{3} Third, the LLM struggles with complex call chains.
Languages like Java introduce deep and polymorphic call hierarchies—unlike Python-based systems where graph search can be bounded—resulting in context truncation or premature termination due to limited context length.

These challenges highlight a fundamental tension: while the knowledge of an entire software project is often too large and unstructured to fit within an LLM’s context window, the lack of explicit domain knowledge makes it difficult to determine which information is truly relevant.
This motivates the use of domain-specific knowledge that can be explicitly extracted, formalized, and selectively exposed to guide LLM reasoning.

In this work, we focus on the concurrency domain—a particularly challenging yet informative subspace of software systems.
Concurrency bugs (e.g., race conditions, deadlocks, atomicity violations) often lack executable test cases and involve subtle cross-thread interactions, making them difficult to localize using existing techniques.
At the same time, concurrency-related knowledge is comparatively structured and bounded.
Execution contexts, shared resources, and access actions form a compact semantic core that can be explicitly identified and formalized.

Based on this observation, we propose \textbf{ConFL}, a novel framework that augments LLM-based fault localization with structured concurrency knowledge.
To address knowledge insufficiency (\blackcircled{1}), ConFL constructs a \emph{Concurrent Knowledge Base (CKB)} that explicitly captures concurrency-relevant entities and behaviors, providing reliable grounding for LLM reasoning.
To mitigate the difficulty of narrowing the search space under incomplete or noisy bug reports (\blackcircled{2}), ConFL employs a \emph{hierarchical retrieval strategy} that progressively localizes faults from components to packages and then to thread entry points.
Finally, to overcome the challenges posed by deep and polymorphic Java call chains (\blackcircled{3}), ConFL represents concurrency behavior using \emph{interaction-level DSL elements}, allowing the LLM to reason directly over shared-resource interactions rather than traversing long call chains.

Through these complementary mechanisms, ConFL bridges the gap between unstructured bug reports and structured concurrency reasoning, enabling more stable, scalable, and explainable fault localization.

We evaluate ConFL on a diverse set of real-world concurrent bugs collected from eight large-scale Java projects.
ConFL achieves 132 Top-1 hits, 186 Top-3 hits, and 216 Top-5 hits, substantially outperforming both traditional IR-based techniques and recent LLM-based baselines.
In terms of ranking quality, ConFL attains an MRR of 0.503 and a MAP of 0.486, improving over the strongest baseline by more than 60\%.
These results demonstrate that explicitly modeling concurrency semantics and performing hierarchy-guided reasoning significantly improves fault localization accuracy under a bug-report–only setting.

In summary, this paper makes the following contributions:
\begin{itemize}
    \item We propose an interaction-level domain-specific language (DSL) that compactly represents concurrency context through thread entries, shared resources, and access actions.
    
    \item We construct a hierarchical Concurrent Knowledge Base (CKB) that organizes structured concurrency knowledge to support controlled and explainable LLM reasoning.
    
    \item We design a hierarchy-guided fault localization framework that progressively narrows the search space and enables interaction-level ranking and explanation.
    
    \item We evaluate our approach on real-world concurrent bugs and show significant improvements in localization accuracy, robustness, and explainability over state-of-the-art baselines.
\end{itemize}

%% file: background.tex
\section{BACKGROUND AND RELATED WORK}

\subsection{Fault Localization}
Traditional fault localization (FL) techniques, such as spectrum-based~\cite{sp1,sp2} and mutation-based~\cite{m1, m2} approaches, rely on dynamic information from test executions. While effective in certain domains, these methods are impractical for concurrent bugs, since constructing reliable and deterministic test cases is notoriously difficult.

To mitigate this dependency, IR-based methods~\cite{yu2008empirical,jones2005empirical,chilimbi2009holmes,santelices2009lightweight, shao2023information} treat bug reports as natural language queries and retrieve candidate source files based on textual similarity. However, these approaches heavily depend on the quality of bug reports. Prior studies have shown that IR-based methods perform poorly at the method level, and are thus typically restricted to file-level localization.

More recently, learning-based approaches~\cite{dl1,dl2} have been proposed. Some methods, such as DeepFL and Grace, still require dynamic coverage or test-case information, while others depend on large-scale training data. Both requirements are unrealistic in the context of concurrency bugs, where labeled data is sparse and projects vary widely in structure. In practice, it remains highly challenging to directly map bug reports to faulty methods.

The emergence of large language models (LLMs) has inspired a new line of research in FL. Systems such as AutoFL~\cite{autofl} and AgentFL~\cite{agentfl} employ interactive prompting to reason over bug reports and test cases, while FlexFL~\cite{flexfl} combines bug report information with test executions. Despite promising results, these methods struggle with fundamental limitations: the lack of comprehensive project understanding, reliance on context-limited interactions, and vulnerability to hallucinations when bug reports are incomplete or misleading.

\subsection{Localizing Concurrency Bugs}

Concurrency bugs are notoriously difficult to localize due to nondeterministic thread scheduling and complex synchronization behavior. Research on concurrency-specific fault localization remains limited. The most closely related work is BLCoiR~\cite{shao2023information}, an IR-based approach that augments bug-report queries with concurrency-related term weighting for file-level retrieval. Unlike ConFL, BLCoiR relies primarily on lexical similarity rather than structured concurrency reasoning.

Another line of work localizes concurrency bugs using dynamic analysis. Approaches such as Falcon~\cite{cf1}, ConSeq~\cite{cf2}, Unicorn~\cite{cf3}, and Codead~\cite{cf4} analyze runtime traces, suspicious interleavings, or memory-access patterns to identify concurrency faults. While effective when failing executions are available, these methods require runtime instrumentation or reproducible executions, which are often difficult to obtain for real-world concurrency bugs.

More broadly, static analyzers such as Chord~\cite{naik2006effective} and RacerD~\cite{blackshear2018racerd} focus on concurrency bug detection rather than fault localization. Although they can identify potential races or deadlocks, they often generate many false positives and do not directly leverage bug reports for localization. In contrast, ConFL localizes concurrency bugs directly from bug reports through hierarchy-guided retrieval and structured interaction reasoning, without requiring execution traces or exhaustive concurrency analysis.

These limitations motivate the need for a lightweight bug-report–driven approach that explicitly models concurrency interactions for fault localization.

%% file: motivation.tex
\section{Problem Statement}

Bug reports are an important source for fault localization.
A high-quality report may directly pinpoint faulty classes or methods; however, in practice, bug reports are often noisy, misleading, or incomplete.
Table~\ref{tab:example} illustrates this challenge using two real-world Apache Druid bugs.

In \textbf{Druid-3174}, the stack trace explicitly mentions \texttt{WorkerTaskMonitor}, yet the actual race condition resides in \texttt{RemoteTaskRunner}, where \texttt{WorkerTaskMonitor} merely serves as a triggering entry point.
Accurate localization therefore requires reasoning beyond the reported class.
In \textbf{Druid-1360}, the bug report contains no explicit class or method names at all, making it particularly difficult for existing fault localization techniques to establish reliable anchors in the codebase.
Without symbolic references, both IR-based and LLM-based approaches struggle to map the report to relevant program locations.

\input{tables/motivation_examples}

We further examine FlexFL, a representative state-of-the-art LLM-based fault localization system that relies solely on bug reports.
FlexFL prompts the LLM to iteratively issue function calls—first resolving class names, then enumerating methods, and finally ranking code snippets.
While effective when bug reports precisely align with program symbols, this workflow degrades significantly under incomplete or misleading reports.
For instance, in Druid-3174, FlexFL remains confined to \texttt{WorkerTaskMonitor}, failing to identify the true fault in \texttt{RemoteTaskRunner}.
In Druid-1360, lacking explicit class names, it incorrectly localizes the fault to \texttt{AutoScaler}, whereas the actual issue lies in \texttt{SimpleResourceManagementStrategy} and \texttt{RemoteTaskRunner}.

These cases highlight three fundamental \textbf{limitations} of bug-report–only LLM-based fault localization for concurrent programs:

\textit{(1) Knowledge insufficiency.}
Bug reports frequently omit critical structural and behavioral information required for grounding LLM reasoning.
When program entities are missing or underspecified, the LLM is forced to infer context from incomplete natural language, leading to unstable or hallucinatory reasoning.
We observe that FlexFL may produce inconsistent localization results across multiple runs on the same report, issuing different function calls and retrieving different candidate methods.

\textit{(2) Ineffective search-space narrowing.}
When bug reports lack explicit program identifiers, LLM-based approaches have no principled way to progressively narrow the search space.
The LLM is therefore forced to issue broad or heuristic queries, often retrieving code only loosely related to the actual fault.
This describes \textbf{Druid-1360}, where the report contains no class or method names and FlexFL incorrectly focuses on \texttt{AutoScaler}, a component only tangentially related to the failure.
Without structured guidance or domain constraints, the LLM fails to identify the relevant concurrency context and repeatedly explores incorrect regions.

\textit{(3) Difficulty reasoning over deep and cross-thread call chains.}
Concurrent Java programs commonly exhibit deep, polymorphic, and cross-thread call hierarchies.
Due to limited context capacity, LLM-based approaches struggle to reason beyond local or downward call expansion.
In \textbf{Druid-3174}, although \texttt{WorkerTaskMonitor} appears in the report, the actual race occurs several layers downstream in \texttt{RemoteTaskRunner}.
Tracing such cross-thread interactions exceeds the practical reasoning depth of current LLM-based pipelines, causing them to remain trapped near the reported entry point.

\noindent
\textbf{Our Solutions.}
To overcome the above limitations, we propose three complementary design choices.

\textit{(1) Structured concurrency knowledge.}
To address knowledge insufficiency, we construct a \emph{Concurrent Knowledge Base (CKB)} that explicitly captures concurrency-relevant entities and behaviors, including thread entry points, shared variables, and synchronization mechanisms.
By grounding LLM reasoning in structured and statically extracted concurrency knowledge, the CKB reduces hallucination and stabilizes reasoning when bug reports lack precise program details.

\textit{(2) Hierarchical search-space narrowing.}
To enable effective localization under incomplete or misleading bug reports, we design a \emph{hierarchical retrieval strategy} that progressively narrows the search space from components to packages and then to thread entry points.
This coarse-to-fine localization process leverages the high cohesion and low coupling of software systems, allowing the LLM to focus on increasingly relevant concurrency contexts while keeping the exposed context bounded.

\textit{(3) Interaction-level reasoning without deep call-chain traversal.}
To overcome the difficulty of reasoning over deep and polymorphic Java call chains, we introduce an \emph{interaction-level DSL} that summarizes concurrency behavior in terms of shared-resource accesses across thread entries.
This representation allows the LLM to reason directly about potential interleavings and synchronization patterns, without explicitly traversing long call chains or enumerating execution paths.

%% file: tables/motivation_examples.tex
\vspace{-0.5em}
\begin{table}[h] \small
\centering
\caption{Bug Reports from Apache Druid} \label{tab:example}
\vspace{-8pt}
\begin{tabularx}{\textwidth}{X}
\Xhline{1.0pt}
\textbf{Title:} Overlord assigns too many tasks to middle manager \textit{\#3174} \\\hline
\textbf{Description:} RTR has a guard against multiple things trying to launch on the same worker via \\
workersWithUnacknowledgedTask.putIfAbsent(immutableZkWorker.get().getWorker().getHost(), ...\\
INFO [WorkerTaskMonitor] io.druid.indexing.worker.WorkerTaskMonitor - Submitting runnable for task ... \\
INFO [WorkerTaskMonitor] io.druid.indexing.worker.WorkerTaskMonitor - Affirmative. Running task ... \\ \hline
\textbf{FlexFL:} 1. \textit{find\_class}("WorkerTaskMonitor") $\rightarrow$ 2. \textit{get\_methods\_of\_class}("WorkerTaskMonitor") $\rightarrow$ 3. \textit{get\_code\_snippet\_of\_method}("mainLoop()") $\rightarrow$ 4. \textit{find\_class}("ForkingTaskRunner") ...\\[5pt]
\textbf{Top 1}: io.druid.indexing.worker.WorkerTaskMonitor.mainLoop() \\
\textbf{Top 2}: io.druid.indexing.overlord.ForkingTaskRunner.run(Task) ... \\
\Xhline{1.0pt}
\textbf{Title:} Race condition in autoscaling terminates nodes that were just assigned tasks \textit{\#1360}\\ \hline
\textbf{Description:} It is possible for the indexing service auto-scaling to terminate nodes that have just been assigned a task, but that have not updated the task status to running yet.
... If you are unlucky and autoscaling checks right after the task has been assigned, but right before the worker announces the task status, and the worker is not running any other tasks, your tasks will get killed. ...\\ \hline
\textbf{FlexFL:} 1. \textit{find\_class}("autoscaling") $\rightarrow$ 2. \textit{get\_methods\_of\_class}("AutoScaler") $\rightarrow$ \\ 3. \textit{get\_code\_snippet\_of\_method}("terminate(List<String>)") ...\\
\textbf{Top 1}: io.druid.indexing.overlord.autoscaling.AutoScaler.terminate(List<String>) \\
\textbf{Top 2}: io.druid.indexing.overlord.autoscaling.AutoScaler.terminateWithIds(List<String>) ... \\
\Xhline{1.0pt}
\end{tabularx}
\end{table}
\vspace{-0.5em}

%% file: approach.tex
\section{Approach}

\begin{figure}
\centering
\includegraphics[width=0.7\textwidth]{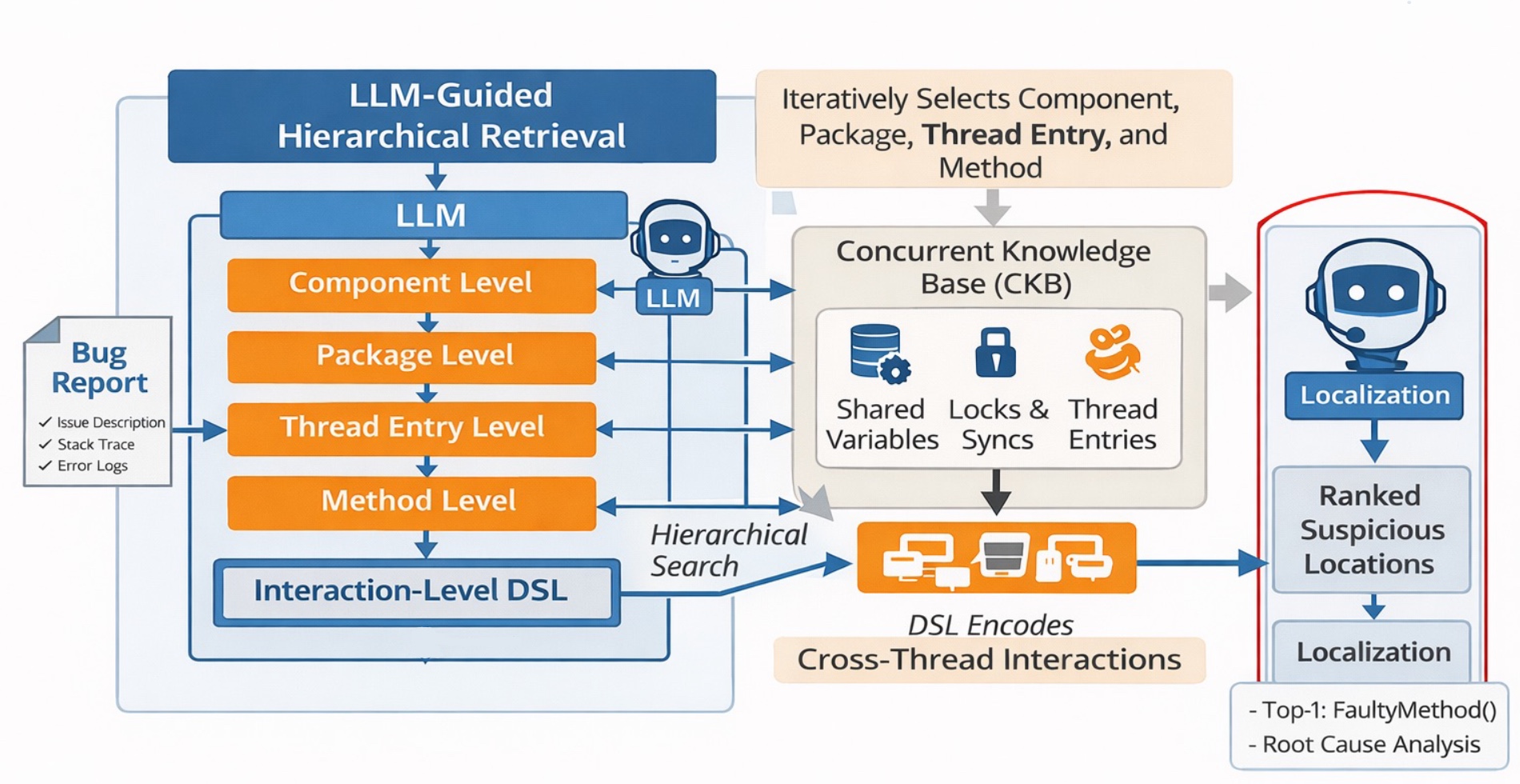}
\vspace{-10pt}
\caption{The overview of \Name{}.} \label{frame}
\vspace*{-10pt}
\end{figure}

As shown in Figure~\ref{frame}, we propose \textbf{ConFL}, 
a hierarchy-guided framework for concurrent fault localization based on structured concurrency knowledge.
Given a bug report as the sole input, ConFL first extracts concurrency-relevant information from source code and organizes it into a Concurrent Knowledge Base (CKB). 
The CKB captures concurrency semantics at multiple abstraction levels, including components, packages, thread entry points, methods, and interaction-level behaviors.

Fault localization is then carried out as a progressive refinement process.
Starting from the bug report, an LLM incrementally queries the CKB to narrow the search space from coarse-grained components to fine-grained concurrency interactions, while keeping the exposed context bounded.
Finally, the LLM reasons over interaction-level DSL representations to rank suspicious methods and generate concurrency-aware explanations.

\input{approach/concurrent_features}

\input{approach/knowlege_base}
\input{approach/fault_localization}

%% file: approach/concurrent_features.tex
\subsection{Interaction DSL Generation}
\label{sec:interaction-dsl}

At the core of our approach is an \emph{interaction DSL} designed to explicitly represent concurrency context in a lightweight and comparable manner across bug reports and source code. The goal of the DSL is to capture concurrency-relevant semantics while filtering out unrelated program details, thereby enabling focused and explainable fault localization.

Recent work has demonstrated the potential of LLMs for fault localization, but a fundamental challenge remains determining what constitutes relevant context. This challenge becomes particularly severe in concurrent systems, where relevant behaviors may span multiple threads, shared resources, and asynchronous execution paths. While general fault localization often requires reasoning over deep call chains and large code regions, we observe that concurrent faults typically exhibit a more structured form of context centered around shared-variable accesses and cross-thread interactions. This motivates a representation that explicitly models concurrency-relevant interactions while suppressing unrelated program information.

\input{tables/entity}

Based on this observation, we propose an interaction DSL that models concurrency at the granularity of shared-variable accesses across thread entry points. For each thread entry, the DSL records shared-variable accesses reachable from its downstream methods, and an interaction is constructed whenever two thread entries may access the same shared resource, whether through the same method or through different methods. Rather than attempting to reconstruct concrete schedules or happens-before relations, the DSL conservatively captures potential cross-thread interactions that may contribute to faults. Each interaction summarizes who may execute concurrently, what resources are shared, where the accesses occur, and the associated synchronization context.

The DSL is intentionally designed as a lightweight program-level abstraction rather than a predictor of bug-specific interactions. The CKB is constructed offline using conservative static analysis to capture methods reachable from thread entry points that access shared variables, which may over-approximate actual runtime behaviors. Instead of treating all extracted interactions as buggy candidates, ConFL further relies on hierarchy-guided retrieval and LLM-based reasoning to identify interactions most relevant to the bug report. This design enables scalable fault localization without requiring execution traces or reproducible thread schedules.

\Name{} targets thread-based concurrency in Java, including explicit \texttt{Thread} creation, \texttt{Executor} submissions, asynchronous callbacks, and framework-managed entry methods. Synchronization is modeled uniformly through both \texttt{synchronized} constructs and \texttt{java.util.concurrent.Lock} primitives. By representing concurrency through shared-variable accesses across thread entry points rather than through language-specific syntax, the DSL naturally supports a broad range of executor-based and asynchronous concurrency patterns commonly used in modern Java systems.

\paragraph{Concurrency Entity Extraction.}
The DSL is grounded in a small set of concurrency-relevant entities, as summarized in Table~\ref{tab:entity-domain}.
All entities correspond to statically available program constructs and can be extracted using lightweight static analysis.
Variables include both class fields and method parameters that may escape their defining scope.
Synchronization via the \texttt{synchronized} keyword is treated separately, as it appears as a method or block modifier, while other synchronization primitives (e.g., \texttt{ReentrantLock}) are modeled as shared variables.
Thread entry points are identified as methods that may initiate concurrent execution, including explicit thread creation, executor submissions, asynchronous callbacks, and framework-managed entry methods.

\begin{wrapfigure}{r}{0.35\textwidth} 
    \centering 
    \includegraphics[width=0.35\textwidth]{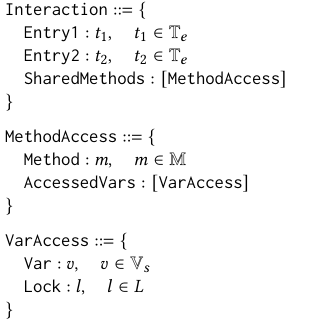}
    \caption{Interaction DSL}\label{fig:interaction-dsl}
\end{wrapfigure}

Based on these extracted entities, we define an interaction DSL to represent potential concurrent behaviors.
Figure~\ref{fig:interaction-dsl} illustrates the structure of the Interaction DSL.
Each interaction is defined by a pair of thread entry points and a set of shared access summaries.
For a given entry pair, the DSL records the methods reachable from each entry that access shared variables, along with the corresponding synchronization context.
By organizing concurrency information at the granularity of entry pairs and access points, the DSL provides a compact yet expressive representation of concurrency-relevant context suitable for subsequent analysis and reasoning.

As shown in Algorithm~\ref{alg:interaction-extraction}, the construction of the Interaction DSL proceeds in three phases.

In the first phase, the algorithm extracts base concurrency-relevant entities from the source code, including methods, shared variables, and thread entry points.
A static call graph is then constructed to support reachability analysis.
These entities form the foundation for subsequent interaction modeling and correspond directly to the analysis domain defined earlier.

In the second phase, the algorithm performs thread-entry–centered reachability analysis.
Starting from each thread entry point, it computes the set of methods that may be executed as part of the corresponding concurrent context.
For each reachable method, the algorithm records all accesses to shared variables.
This step associates downstream shared accesses with their originating thread entries, enabling the DSL to capture cross-class and cross-call concurrency behavior.

In the final phase, the algorithm constructs interaction instances by examining pairs of thread entry points.
For each entry pair, it identifies methods that are reachable from both entries and aggregates their shared variable accesses.
An interaction is created whenever two thread entries may access shared state through one or more common methods.
Each interaction summarizes the potential concurrent behavior between the two entries by grouping the relevant methods and shared accesses into a single DSL element.

Overall, the algorithm conservatively over-approximates concurrency interactions: it captures all shared accesses that may occur under some interleaving, without committing to specific execution orders or schedules.
This design ensures soundness while keeping the interaction representation compact and suitable for scalable fault localization and LLM-based reasoning.

\vspace{-0.5em}
{\footnotesize
\begin{algorithm}[h]
\caption{Interaction DSL Extraction}
\label{alg:interaction-extraction}
\begin{algorithmic}[1]
\Require Source code $\mathcal{P}$
\Ensure Set of interactions $\mathcal{I}$

\State \textbf{// Phase 1: Extract base entities}
\State $\mathbb{M} \gets \textit{ExtractMethods}(\mathcal{P})$
\State $\mathbb{V}_s \gets \textit{ExtractVars}(\mathcal{P})$
\State $\mathbb{T}_e \gets \textit{IdentifyThreadEntries}(\mathbb{M})$
\State $\mathcal{G} \gets \textit{BuildCallGraph}(\mathbb{M})$

\State \textbf{// Phase 2: Compute reachability and accesses per thread entry}
\ForAll{$t \in \mathbb{T}_e$}
    \State $\textit{Reach}(t) \gets \textit{ComputeReachable}(\mathcal{G}, t.m)$
    \ForAll{$m \in \textit{Reach}(t)$}
        \State $\textit{Accesses}(t, m) \gets \textit{ExtractAccesses}(m, \mathbb{V}_s)$
    \EndFor
\EndFor

\State \textbf{// Phase 3: Build interactions for each thread entry pair}
\State $\mathcal{I} \gets \emptyset$
\ForAll{$(t_1, t_2) \in \mathbb{T}_e \times \mathbb{T}_e$ where $t_1 < t_2$}
    \State $M_{common} \gets \textit{Reach}(t_1) \cap \textit{Reach}(t_2)$
    \State $\textit{SharedMethods} \gets \emptyset$
    \ForAll{$m \in M_{common}$}
        \State $A_m \gets \textit{Accesses}(t_1, m) \cup \textit{Accesses}(t_2, m)$
        \If{$A_m \neq \emptyset$}
            \State $\textit{SharedMethods} \gets \textit{SharedMethods} \cup \{(m, A_m)\}$
        \EndIf
    \EndFor
    \If{$\textit{SharedMethods} \neq \emptyset$}
        \State $I \gets \textit{BuildInteraction}(t_1, t_2, \textit{SharedMethods})$
        \State $\mathcal{I} \gets \mathcal{I} \cup \{I\}$
    \EndIf
\EndFor

\State \Return $\mathcal{I}$
\end{algorithmic}
\end{algorithm}
}
\vspace{-0.5em}

%% file: tables/entity.tex
\vspace{-0.5em}
\begin{table}[H]
\centering \scriptsize
\caption{Analysis Domain for Interaction DSL}
\vspace{-8pt}
\label{tab:entity-domain}
\renewcommand{\arraystretch}{1.25}
\begin{tabular}{@{}l l l@{}}
\toprule
\textbf{Entity} & \textbf{Symbol} & \textbf{Definition} \\
\midrule
method
  & $m \in \mathbb{M}$ 
  &  \\

variable
  & $v \in \mathbb{V}_{s}$ 
  &  $::= Field \cup Param$\\

lock 
  & $l \in \mathbb{L}$ 
  & $::= synchronized$ \\

thread entry point
  & $t \in \mathbb{T}_e$ 
  & $\subseteq \mathbb{M}$  \\
\bottomrule
\end{tabular}
\end{table}
\vspace{-8pt}

%% file: approach/knowlege_base.tex
\subsection{Concurrent Knowledge Base}
\label{sec:ckb}

Based on the extracted concurrency entities and interaction DSL described in the previous section, we construct a \emph{Concurrent Knowledge Base (CKB)} to organize concurrency-relevant information in a hierarchical and queryable form.
The CKB serves as the structured knowledge layer that supports hierarchical retrieval and LLM-guided fault localization.

Formally, the Concurrent Knowledge Base is defined as:
\[
\mathcal{K} = \{ \mathcal{C}, \mathcal{P}, \mathcal{E}, \mathcal{M}, \mathcal{I} \}
\]
where each element captures concurrency knowledge at a different abstraction level.

The Concurrent Knowledge Base (CKB) organizes concurrency-relevant information at multiple abstraction levels to support progressive fault localization.
At coarse granularity, component- and package-level knowledge captures high-level system structure and functionality.
Each component and package is associated with brief descriptions summarizing its responsibilities and concurrency relevance, derived from build configurations, directory structure, and developer documentation.
This information provides early-stage guidance for narrowing the search space when analyzing large codebases.

At finer granularity, the CKB records detailed concurrency semantics at the thread-entry, method, and interaction levels.
Formally, thread-entry knowledge corresponds to the set $\mathbb{T}_e$, where each $t \in \mathbb{T}_e$ represents a concurrency entry point such as a thread, executor submission, or asynchronous callback.
For each thread entry $t$, the CKB records the set of other entries $t' \in \mathbb{T}_e$ with which $t$ may interact through shared resource accesses.
Method-level knowledge is defined over the set of methods $\mathbb{M}$.
For each method $m \in \mathbb{M}$, the CKB records whether $m$ serves as a thread entry ($m \in \mathbb{T}_e$), the set of shared variables $v \in \mathbb{V}_s$ accessed by $m$, and the corresponding access types (e.g., read or write).
This information captures how concurrency behavior manifests at the code level.
Finally, interaction-level knowledge consists of interaction DSL elements $\mathcal{I}$.
Each interaction $I \in \mathcal{I}$ summarizes a potential concurrent behavior between a pair of distinct thread entries $(t_1, t_2)$, where $t_1, t_2 \in \mathbb{T}_e$.

\vspace{-0.5em}
\input{tables/fun_call}
\vspace{-0.5em}

\paragraph{Retrieval Functions}
\label{subsec:ckb-retrieval}

To enable hierarchical and controlled access to the Concurrent Knowledge Base, we expose a small set of retrieval functions, summarized in Table~\ref{tab:function_calls}.
Each function corresponds to a specific abstraction level in the knowledge base, ranging from components and packages to thread entries, methods, and interaction DSL elements.
By restricting retrieval to these predefined functions, the framework ensures that the LLM incrementally explores only concurrency-relevant context, preventing uncontrolled traversal of the codebase and keeping the reasoning process focused and interpretable.

%% file: tables/fun_call.tex
\begin{table}[h]
\centering
\caption{Function calls for hierarchical retrieval}
\vspace{-8pt}
\label{tab:function_calls}
\scriptsize
\begin{tabular}{l|l|p{7.5cm}}
\hline
\textbf{Name} & \textbf{Arguments} & \textbf{Description} \\
\hline
list\_components & None &
Return all components with brief descriptions. \\
\hline
list\_packages\_under\_component & component\_name &
Return all packages under the specified component. \\
\hline
list\_entries\_under\_package & package\_name &
Return all thread entry points defined in the package. \\
\hline
list\_method\_under\_class & class\_name &
Return all methods under the specified class. \\
\hline
get\_entry\_summary & entry\_name &
Return concurrency summary of the thread entry and its interacting entries. \\
\hline
get\_method\_summary & method\_name &
Return concurrency-related information of the method. \\
\hline
get\_interaction\_dsl & entry\_pair &
Return interaction DSL for the given pair of thread entries. \\
\hline
get\_code\_snippet & method\_name &
Return code snippet of the specified method. \\
\hline
exit & None &
Terminate the retrieval process. \\
\hline
\end{tabular}
\end{table}
\vspace{-8pt}

%% file: approach/fault_localization.tex
\subsection{Explainable Fault Localization}
\label{sec:explainable-fl}

Given a bug report, our fault localization procedure follows a \emph{progressive refinement} strategy over the Concurrent Knowledge Base (CKB).
Rather than exposing large code fragments or entire program graphs to an LLM, we localize concurrency faults by performing hierarchical retrieval and reasoning over structured, DSL-encoded concurrency knowledge.

The hierarchical design is motivated by the structural properties of large software systems.
Well-engineered systems are typically organized with high cohesion and low coupling: components differ significantly in functionality, packages encapsulate closely related behaviors, and thread entry points define distinct concurrent execution contexts.
As a result, concurrency faults are usually confined to a small region of the system and rarely span arbitrary components or unrelated execution contexts.
This observation motivates a coarse-to-fine localization strategy that progressively narrows the search space along natural architectural boundaries.

\paragraph{Hierarchical Localization Workflow.}
Starting from the bug report, the LLM incrementally refines the localization scope by querying the CKB in the following order:
(i) selecting the most relevant component,
(ii) selecting a package within the chosen component,
(iii) localizing candidate thread entry points within the package, and
(iv) selecting a pair of interacting thread entries.
At each stage, only a small and relevant slice of the CKB is exposed to the LLM (e.g., a list of components, packages, or thread entries), ensuring that the reasoning context remains bounded and focused.

Once a thread-entry pair is selected, the framework retrieves all interaction-level records associated with the pair.
These records are represented using the interaction DSL and summarize shared resources, access actions, and synchronization context when available.
The interaction set provides a compact, concurrency-focused context that identifies the candidate methods involved in the interaction.

\paragraph{LLM-based Interaction Ranking and Explanation.}
Given the interaction DSL elements, the LLM performs interaction-level reasoning to assess which methods are most likely responsible for the fault.
To support this reasoning, the LLM may selectively query method-level summaries and, when necessary, corresponding code snippets through predefined retrieval functions.
Based on the combined interaction context and method-level information, the LLM produces a ranked list of suspicious methods.

Finally, the LLM generates an explanation grounded in explicit concurrency context.
The explanation links the ranked methods to their accessed shared variables and the corresponding thread-entry pair, describing how concurrent execution and shared-resource access may lead to the observed faulty behavior.

\begin{figure}[H] \scriptsize
\centering 
\vspace*{3pt}
\begin{tcolorbox}[colback=gray!5,colframe=black!30]

You are an expert in concurrent fault localization. Given a bug report and access to a Concurrent Knowledge Base (CKB), iteratively retrieve relevant concurrency information through hierarchical reasoning and identify the most suspicious methods related to the reported fault.

\vspace{0.3em}

\textbf{Bug Report:} \verb!{{BUG_REPORT}}!

\# Component-Level Retrieval

First, call: \textit{$list\_components()$}. Use function calls to retrieve concurrency-related knowledge from the CKB, then analyze the returned components and identify the most relevant component associated with the bug report.

\begin{verbatim}
{{function_call}}
\end{verbatim}
\textbf{Output Format:} component\_name: component
\vspace{0.3em}

\# Package-Level Retrieval

First, call: \textit{$list\_packages\_under\_component(component)$}. Use function calls to retrieve concurrency-related knowledge from the CKB, then analyze the returned packages and identify the most relevant package associated with the bug report. 

\begin{verbatim}
{{function_call}}
\end{verbatim}
\textbf{Output Format:} package\_name: package
\vspace{0.3em}

\# Thread Entry and Method-Level Retrieval

First, call: \textit{$list\_entries\_under\_package(package)$}. Use function calls to retrieve concurrency-related knowledge from the CKB, then analyze the returned thread entry points and identify the most suspicious entry pair related to the reported fault. 

Select the most suspicious thread-entry pair.

\begin{verbatim}
{{function_call}}
\end{verbatim}
\textbf{Output Format:} entry\_pair: (entry1, entry2)
\vspace{0.3em}

\# Interaction DSL Reasoning

First, retrieve the interaction DSL: $get\_interaction\_dsl(entry_pair)$. Use the DSL to analyze shared resources, access actions, synchronization context, and cross-thread interactions. If more evidence is needed, continue using function calls to inspect relevant method summaries or source code:
\begin{verbatim}
{{function_call}}
\end{verbatim}

After sufficient evidence has been collected, apply the reasoning rules to identify suspicious methods involved in the concurrent interaction. Then return a ranked list of suspicious methods and explain how the interaction may lead to the reported fault.

\textbf{Output Format:}

\begin{verbatim}
Rank 1: [Method]: ... [Shared Resource]: ... [Thread Entries]: ... [Explanation]: ...
\end{verbatim}
\end{tcolorbox}
\vspace{-0.4em}
\caption{Prompt Template} \label{template}
\vspace{-0.15in}
\end{figure}

\paragraph{Prompt Template.} 
ConFL follows the function-calling paradigm common to recent LLM-based FL approaches (e.g., AutoFL, FlexFL). Figure~\ref{template} illustrates the complete prompt template. At each hierarchical retrieval stage, the LLM is provided with the bug report and a typed set of retrieval functions (listed in Table~\ref{tab:function_calls}); it selects one function to invoke, observes the structured response, and progresses to the next stage. This constrains the LLM to operate within the CKB schema rather than generating free-form responses. At the interaction-ranking stage, each interaction DSL element is serialized as a labeled block listing the entry pair, the set of shared methods, and the per-method shared-variable accesses with their guard locks, so that the LLM can reason directly over concurrency-relevant context without traversing raw code. We use deterministic decoding (temperature = 0) and the complete prompt templates are released in the artifact.

%% file: evaluation.tex
\section{Evaluation}

We will answer the following research questions.  

\vspace*{3pt}
\noindent
{\bf RQ1: How effective is our approach in localizing concurrent bugs?}

\vspace*{3pt}
\noindent
{\bf RQ2: How does ConFL compare with state-of-the-art across different bug-report conditions?}

\vspace*{3pt}
\noindent
{\bf RQ3: What is the contribution of each component in our approach?}

\vspace*{3pt}
\noindent
{\bf RQ4:  How well does our approach generalize to unseen data and different LLMs?}

\subsection{Experimental Setup}

\subsubsection{Dataset}

The first dataset is derived from projects used in prior concurrent fault localization~\cite{shao2023information} and bug report classification studies~\cite{shao2026identifying}. As shown in Table~\ref{table:git1}, Dataset$_{Git}$ contains eight large-scale open-source systems, including druid, grpc-java, presto, pulsar, redisson, rocketmq, trino, and vert.x, with 4--100 concurrency bug reports per project. The projects range from roughly 9K to over 700K methods and cover diverse concurrency styles, including lock-based synchronization, executor-based concurrency, asynchronous callbacks, and mixed interaction patterns. Ground-truth faulty methods are identified from bug-fixing commits.

To evaluate generalization beyond potential training-data exposure, we additionally construct a post-cutoff dataset containing concurrency bugs reported after the LLM training cutoff date. We follow the same collection procedure as Dataset$_{Git}$ and retain only bugs with concurrency-related descriptions, associated fix commits, and verifiable faulty methods. Since several Dataset$_{Git}$ projects yielded few eligible bugs, we further include widely used Java systems frequently used in concurrency-related research, including Camel, Cassandra, Hadoop, and HBase~\cite{shao2023information, shao2026identifying}. As shown in Table~\ref{table:git2}, the resulting dataset contains 32 bugs in total.

\input{tables/results/dataset}

\subsubsection{Baseline Selection.}

Our framework operates strictly in a bug-report–only setting. Therefore, we select baselines that can be executed under the same constraint and that represent the strongest available techniques in their respective categories.

\noindent
\textbf{IR-Based Baselines.}
Most IR-based FL techniques work at the file level and are not directly applicable to method-level localization. We include two representative methods: \textbf{BRTracer} \cite{BTrace} — shown in prior studies \cite{shao2024enhancing, lee2018bench4bl} to be one of the strongest IRFL baselines. We adapt it to method-level by treating each method as an independent document. \textbf{BoostNSift} \cite{razzaq2021boostnsift} — a recent IR-based method designed specifically for method-level FL, providing a competitive IR baseline for our evaluation. These baselines represent the upper bound of lexical-similarity approaches under bug-report–only conditions. We additionally include \textbf{BLCoiR}~\cite{shao2023information}, the only prior bug-report–based fault localization technique specifically designed for concurrent bugs. Since BLCoiR originally operates at file-level granularity, we similarly adapt it to method-level retrieval by treating each method as an independent document. For file-level comparison, we aggregate \Name{}'s method-level outputs using a max-score rule, where the score of a file is defined as the maximum score among its methods. Together, these baselines represent strong lexical-similarity approaches under bug-report–only settings.

\noindent
\textbf{LLM-Based Baselines.}
Many recent LLM-based approaches (e.g., AutoFL, AgentFL) rely on test-case execution and cannot run without dynamic information. As discussed in our motivation, concurrent bugs rarely have reliable test cases, making these methods incompatible with our evaluation setting.
We therefore select: \textbf{FlexFL} — a recent LLM-based FL framework that supports bug-report–only mode and thus can be fairly compared with ours.

\noindent
\textbf{Other Baselines.}
Supervised FL techniques such as DreamLoc~\cite{dreamloc} and rLocator~\cite{locator} require thousands of labeled bugs per project, while our projects contain only 4–40 real-world concurrent bugs, making such training infeasible.
Other learning-based methods like DeepFL depend on mutation-based synthetic bugs and large test suites, which are unavailable for concurrent bug scenarios. Thus, these approaches cannot serve as practical baselines.

Recent LLM  methods for Python (e.g., CoSIL) rely on building full program graphs and performing LLM-guided exploration. These techniques do not scale to Java due to: call graphs with hundreds of thousands to millions of nodes,	extensive virtual dispatch, and lack of existing Java adaptations.

Spectrum-based and hybrid techniques (e.g., Tarantula~\cite{Tarantula}, Ochiai~\cite{abreu2009practical}, DStar~\cite{dst}, DeepFL~\cite{dl1}, Grace~\cite{dl2}) rely on failing/passing tests and coverage information, which are difficult to obtain reliably for concurrency bugs due to nondeterministic scheduling. A concurrency test may pass or fail across runs without code changes, weakening coverage-based statistical correlation. Supervised learning-based methods also require large amounts of labeled training data, which is impractical given the limited number of concurrency bugs available per project in our benchmark (4--100 bugs/project).

\subsubsection{Performance Metrics.}

We evaluate \Name{} using three standard metrics widely adopted in prior fault localization studies~\cite{flexfl, autofl,yu2008empirical,jones2005empirical,chilimbi2009holmes,santelices2009lightweight}.

\textbf{Top@K} measures the proportion of bug reports whose faulty method appears within the top-$k$ results ($k\!\in\!\{1,5,10\}$), defined as
$Top@K = |R_k| / n$, where $|R_k|$ is the number of reports with at least one correct faulty method ranked in the top-$k$, and $n$ is the total number of bug reports.

\textbf{MRR} (Mean Reciprocal Rank) captures how early the first faulty method appears:
$MRR = \frac{1}{n}\sum_{j=1}^{n} \frac{1}{rank_j}$,
where $rank_j$ is the rank of the first relevant faulty method for the $j$-th report.

\textbf{MAP} (Mean Average Precision) evaluates the overall ranking quality of all faulty methods:
$MAP = \frac{1}{n}\sum_{j=1}^{n} AvgP_j$,
where $AvgP_j = \frac{1}{|K_j|}\sum_{k\in K_j} Prec@k$ and
$Prec@k = \frac{1}{k}\sum_{i=1}^{k} IsRelevant(i)$.
Here, $K_j$ denotes the set of true faulty methods for report $j$, and $IsRelevant(i)=1$ if the $i$-th ranked method is faulty. Higher values indicate better localization performance.

\subsection{Results Analysis}

\input{RQs/RQ1}

\input{RQs/RQ2}

\input{RQs/RQ3}
\input{RQs/RQ4}

%% file: tables/results/dataset.tex
\begin{table*}[t]\footnotesize
\centering
\begin{minipage}{0.48\linewidth}
\centering
\caption{Dataset$_{Git}$}
\label{table:git1}
\begin{tabular}{c|ccc}
\toprule
Project & \#BR & Avg.\#M & Avg.\#F \\ \hline
druid  & 31      & 22,692      &  2,954        \\ 
grpc-java    &  43     & 9,970      & 643       \\ 
presto    &  7     &  23,350     &  2,790         \\ 
pulsar    &  72     &  27,917     & 2,998          \\ 
redisson    & 100       &  14,262     & 1,216          \\ 
rocketmq    & 4      & 15,123      & 1,589         \\ 
trino    & 17      &  708,356     & 8,160          \\ 
vert.x     & 48      &  11,525     & 761    \\ 
\textbf{total}     & 322      &  14,251  &  1,808   \\ 
\bottomrule
\end{tabular}
\end{minipage}
\hfill
\begin{minipage}{0.48\linewidth}
\centering
\caption{Dataset$_{Post}$}
\label{table:git2}
\begin{tabular}{c|ccc}
\toprule
Project & \#BR & Avg.\#M & Avg.\#F \\ \hline
druid  & 1      & 73,782      & 8,608         \\ 
grpc-java    & 4      & 20,839      & 1,512       \\ 
pulsar    & 13      & 37,971      &  4,200         \\ 
redisson    & 3      & 31,777      & 2,609          \\ 
camel    & 4      & 180,000      & 22,406          \\ 
cassandra    & 2      & 42,408     & 3,463          \\ 
hadoop     & 1      & 136,170      & 11,829    \\ 
hbase     & 4      &  69,600     & 5,035    \\ 
\textbf{total}    & 32      & 72,945      &  6,425   \\ 
\bottomrule
\end{tabular}
\end{minipage}
\end{table*}
\vspace{-0.5em}

%% file: RQs/RQ1.tex
\subsubsection{RQ1: How effective is our approach in localizing concurrent bugs?}

Table~\ref{table:overall} presents the fault localization performance of our approach on DatasetGit.
ConFL achieves 132 Top-1 hits, 186 Top-3 hits, and 216 Top-5 hits, substantially outperforming all baselines.
In contrast, the strongest baseline, FlexFL, achieves 82 Top-1 hits, 110 Top-3 hits, and 123 Top-5 hits, while IR-based techniques perform significantly worse.

In terms of ranking quality, ConFL attains an MRR of 0.503 and a MAP of 0.486, improving upon FlexFL by +0.189 MRR and +0.178 MAP.
Compared to traditional IR-based approaches (BRTracer and BoostNSift), the improvements are even more pronounced, exceeding +0.40 in both MRR and MAP.
These results indicate that ConFL not only identifies the correct faulty methods more frequently, but also ranks them more consistently near the top of the candidate list.

\input{tables/results/overall_results}

\noindent
\textbf{Comparison with BLCoiR.} BLCoiR performs poorly at the method level, achieving only 0.030 MRR and 0.030 MAP. This is mainly due to a granularity mismatch between its sparse, file-level query design and method-level retrieval. BLCoiR retains only coarse-grained concurrency keywords and class-level information, which are effective for file-level retrieval where files contain rich aggregated context, but insufficient for individual methods that are themselves sparse and localized. To provide a fairer comparison, we additionally evaluate both techniques at file-level granularity, following the original BLCoiR setting. As shown in Table~\ref{tab:file}, \Name{} still substantially outperforms BLCoiR, improving MRR from 0.32 to 0.76 and MAP from 0.21 to 0.70. These results suggest that \Name{}'s gains primarily come from structured concurrency interaction modeling rather than localization granularity alone.

\begin{wraptable}{r}{0.3\textwidth} \footnotesize
\vspace{-10pt}
\centering 
\caption{File-level comparison.}
\vspace{-8pt}
\label{tab:file}
\begin{tabular}{lcc}
\toprule
Technique & MRR & MAP \\
\midrule
BLCoiR & 0.32 & 0.21 \\
ConFL & \textbf{0.76} & \textbf{0.70} \\
\bottomrule
\end{tabular}
\vspace{-10pt}
\end{wraptable}

Overall, the strong gains in Top-$k$ accuracy, MRR, and MAP on Dataset$_{Git}$ demonstrate the effectiveness of ConFL in localizing real-world concurrent bugs using bug reports alone, even in large and diverse codebases mined from GitHub.

\noindent\textbf{Evaluating Explanation Quality.}
As described in Section~4.3, ConFL explanations identify suspicious methods, thread entry points, shared resources, and synchronization context. The goal of this evaluation is to assess whether these explanations faithfully describe the concurrency behavior of the localized fault rather than their fluency or writing quality.

\noindent\textbf{Evaluation Setup.}
Automated metrics such as BLEU or BERTScore cannot determine whether claims about thread interactions are factually correct, and no benchmark of ground-truth concurrency explanations exists. We therefore conduct manual evaluation against the source code and developer fix commits. Annotators classify each explanation as \textbf{Faithful} (all concurrency claims are supported by the code), \textbf{Partial} (the explanation is generally correct but omits relevant interactions), or \textbf{Hallucinated} (the explanation contains unsupported thread interactions or synchronization behavior).

We evaluate only successfully localized bugs, since explanations based on incorrect localizations are not meaningful. We randomly sample 50 bugs and generate three explanations per bug under two settings: \textit{w/ DSL}, where the LLM receives interaction DSLs, and \textit{w/o DSL}, where the LLM receives only raw source code. This yields 300 explanations in total. Two annotators independently labeled all explanations, achieving 88.3\% agreement with Cohen's $\kappa = 0.767$, indicating substantial agreement.

\begin{wraptable}{r}{0.32\textwidth}\footnotesize
\vspace{-10pt}
\centering
\caption{Explanation quality.}
\vspace{-8pt}
\label{tab:explanation}
\begin{tabular}{lcc}
\toprule
Category & w/ DSL & w/o DSL \\
\midrule
Faithful     & \textbf{78\%} & 70\%          \\
Partial      & 10\%          & 5\%           \\
Hallucinated & \textbf{12\%} & 25\%          \\
\bottomrule
\end{tabular}
\vspace{-10pt}
\end{wraptable}

\noindent\textbf{Results.}
Table~\ref{tab:explanation} shows that the interaction DSL improves the faithful rate from 70\% to 78\% while reducing hallucinations from 25\% to 12\%. This reduction is particularly important because hallucinated thread interactions or synchronization behavior can mislead developers during debugging. The improvement stems from the DSL grounding explanations in explicit thread entry points, shared variables, and access relationships extracted from static analysis, thereby constraining the LLM to reason over verifiable concurrency structure.

\noindent\textbf{Illustrative Case.}
The motivating example \textbf{Druid-3174} illustrates this effect. Without the DSL, the LLM explains only local behavior inside \texttt{RemoteTaskRunner}. With the DSL, ConFL additionally captures that \texttt{WorkerTaskMonitor} initiates concurrent execution while \texttt{RemoteTaskRunner} launches additional threads accessing shared task-state resources, enabling the explanation to describe the fault in terms of concrete cross-thread interactions.

\vspace{-0.5em}
\vspace*{3pt}
\begin{tcolorbox}
\stepcounter{findingsCounter}
{\bf RQ\thefindingsCounter{} summary:} {
The results show that ConFL significantly outperforms all baselines on Dataset$_{Git}$, achieving higher Top-$k$ accuracy as well as substantially better MRR and MAP.
These gains indicate that ConFL not only localizes faulty methods more effectively, but also ranks them more reliably near the top.
Moreover, by providing interaction-level concurrency context, ConFL enables more informative explanations than method-level outputs alone, supporting both accurate and explainable fault localization using bug reports only.
}
\end{tcolorbox}

%% file: tables/results/overall_results.tex
\vspace{-8pt}
\begin{table}[H]\footnotesize
\caption{Performance on Dataset$_{Git}$} \label{table:overall}
\vspace{-8pt}
\begin{tabular}{c|ccc|cc}
\toprule
Technique & Top-1 & Top-3 & Top-5 & MRR & MAP \\ \hline
BRTracer  &  18     &  32     &  47     & 0.084    & 0.082     \\ 
BoostN    &   24    &  41     &   50    & 0.103    &  0.099   \\ 
BLCoiR  &  5     &  11    &  20     & 0.030    & 0.030     \\ \hline
Flexfl    &  82     &  110     &  123     &  0.314   & 0.308    \\ 
ConFL     &  132     &  186     &  216     & \textbf{0.503}    & \textbf{0.486}    \\ \bottomrule
\end{tabular}
\end{table}
\vspace{-8pt}

%% file: RQs/RQ2.tex
\subsubsection{RQ2: How does ConFL compare with state-of-the-art across different bug-report conditions?}

While RQ1 establishes that ConFL outperforms both IR-based baselines (BRTracer, BoostNSift) and the LLM-based baseline FlexFL on aggregate metrics, it does not reveal where this improvement comes from or under what conditions each technique succeeds or fails. To identify the source of ConFL's gains, we stratify the evaluation by bug-report quality, comparing all techniques under three reporting conditions that reflect different levels of structural guidance, following prior fault-localization studies~\cite{shao2024enhancing, blizzard, shao2023information}.
Specifically, we consider: (i) PE reports that include explicit identifiers aligned with the ground truth (PE with hint),
(ii) PE reports that contain misleading or irrelevant identifiers (PE without hint), and
(iii) NL-only reports that provide only natural-language descriptions without code identifiers.

Table~\ref{tab:db} summarizes the results.
When strong structural hints are present (PE with hint), all LLM-based approaches benefit substantially.
In this setting, ConFL achieves the best performance (MRR = 0.756, MAP = 0.724), outperforming FlexFL by +0.15 MRR and +0.13 MAP, and significantly surpassing IR-based baselines.
This result indicates that even with accurate identifiers, concurrency-aware static context provides complementary signals beyond surface-level matching.

The \emph{PE without hint} setting is the most challenging, as misleading program-entity mentions can actively steer LLM reasoning toward incorrect code regions.
Under this condition, FlexFL suffers a severe performance collapse (MRR = 0.021, MAP = 0.021), performing comparably to or worse than IR-based methods.
In contrast, ConFL degrades more gracefully, retaining substantially higher performance (MRR = 0.189, MAP = 0.189).
Although noisy identifiers may still introduce false positives, hierarchy-based retrieval prevents the LLM from overfitting to individual symbols and enables more stable localization by grounding reasoning in concurrency-relevant structure.

Interestingly, NL-only reports are often easier to handle than \emph{PE without hint} reports.
While NL-only reports provide less explicit information, they do not actively mislead the reasoning process.
As a result, ConFL achieves higher performance on NL-only reports (MRR = 0.320, MAP = 0.321) than on PE without hint reports (MRR = 0.189, MAP = 0.189), as hierarchy-based retrieval can explore concurrency-relevant components and thread entries in a more unbiased manner.

\vspace{-8pt}
\input{tables/results/results_by_types}

\vspace{-8pt}

Overall, these results show that although LLM-based techniques can perform well when guided by accurate program-entity hints, they are highly sensitive to noise and misleading information.
By explicitly modeling concurrency interactions and decoupling localization from unreliable lexical signals, ConFL achieves both superior accuracy and improved robustness under realistic and noisy bug-report conditions.

\noindent
\textbf{Case Study.} We revisit the motivating example druid-1360, to briefly illustrate how our framework performs hierarchical fault localization and produces explainable results in practice.
Starting from the bug report, the framework first identifies the \texttt{indexing-service} component and further narrows the search space to the package \texttt{io.druid.indexing.overlord.autoscaling}.
Within this scope, \texttt{SimpleResourceManagement} is identified as a key thread entry point involved in autoscaling decisions, which interacts with other concurrent execution contexts such as the \texttt{RemoteTaskRunner}.

By constructing interaction-level DSL representations for these thread entries, the framework exposes concurrent accesses to shared task and worker state.
Based on this interaction-level context, the LLM identifies a race between task provisioning and termination, localizing the fault to \texttt{doProvision} and \texttt{doTerminate} in \texttt{SimpleResourceManagement}.
This example demonstrates how hierarchical retrieval and interaction-level modeling guide the LLM to focus on a small set of concurrency-critical methods, enabling concise and concurrency-aware explanations without exploring the entire codebase.

\vspace{-0.5em}
\vspace*{3pt}
\begin{tcolorbox}
\stepcounter{findingsCounter}
{\bf RQ\thefindingsCounter{} summary:} {
Across all bug-report conditions, ConFL consistently outperforms both IR-based and LLM-based baselines.
While LLM-based methods benefit from accurate program-entity hints, they are highly sensitive to noisy or misleading identifiers.
By grounding reasoning in structured concurrency knowledge and hierarchy-based retrieval, ConFL achieves superior robustness and maintains strong localization performance even under misleading or identifier-free bug reports.
These results demonstrate that explicitly modeling concurrency interactions is critical for reliable fault localization in realistic settings.
}
\end{tcolorbox}

%% file: tables/results/results_by_types.tex
\begin{table*}[h]
\centering \footnotesize
\caption{MRR and MAP results under different bug report qualities.}\label{tab:db}
\vspace{-8pt}
\begin{subtable}{0.32\textwidth}
\centering
\caption{PE (with hint)}
\begin{tabular}{lcc}
\toprule
Technique & MRR & MAP \\
\midrule
BRTracer & 0.136 & 0.134 \\
boostNSift & 0.167 & 0.161 \\
BLCoiR & 0.048 & 0.048 \\
FlexFL & 0.607 & 0.597 \\
\Name{} & \textbf{0.756} & \textbf{0.724} \\
\bottomrule
\end{tabular}
\end{subtable}
\hfill
\begin{subtable}{0.32\textwidth}
\centering
\caption{PE (without hint)}
\begin{tabular}{lcc}
\toprule
Technique & MRR & MAP \\
\midrule
BRTracer & 0.026 & 0.024 \\
boostNSift & 0.042 & 0.040 \\
BLCoiR & 0.007 & 0.007 \\
FlexFL & 0.021 & 0.021 \\
\Name{} & \textbf{0.189} & \textbf{0.189} \\
\bottomrule
\end{tabular}
\end{subtable}
\hfill
\begin{subtable}{0.32\textwidth}
\centering
\caption{NL}
\begin{tabular}{lcc}
\toprule
Technique & MRR & MAP \\
\midrule
BRTracer & 0.052 & 0.054 \\
boostNSift & 0.054 & 0.049 \\
BLCoiR & 0.018 & 0.020 \\
FlexFL & 0.023 & 0.021 \\
\Name{} & \textbf{0.320} & \textbf{0.321} \\
\bottomrule
\end{tabular}
\end{subtable}
\end{table*}

%% file: RQs/RQ3.tex
\subsubsection{RQ3: What is the contribution of each component in our framework?}

To quantify the contribution of each major component in ConFL, we conduct an ablation study by disabling one component at a time.
Specifically, we evaluate three variants:
(i) \textbf{w/o Hierarchy}, which removes hierarchical retrieval and directly performs flat retrieval over the entire search space;
(ii) \textbf{w/o CKB}, which disables the concurrency knowledge base and provides only basic method-level information without structured concurrency semantics; and
(iii) \textbf{w/o DSL}, which removes the interaction-level DSL representation and feeds unstructured concurrency information to the LLM.
The full system (\textbf{ConFL}) serves as the reference.

\vspace{-0.5em}
\input{tables/results/ablation}

\begin{wrapfigure}{r}{0.35\textwidth} 
    \centering 
    \includegraphics[width=0.35\textwidth]{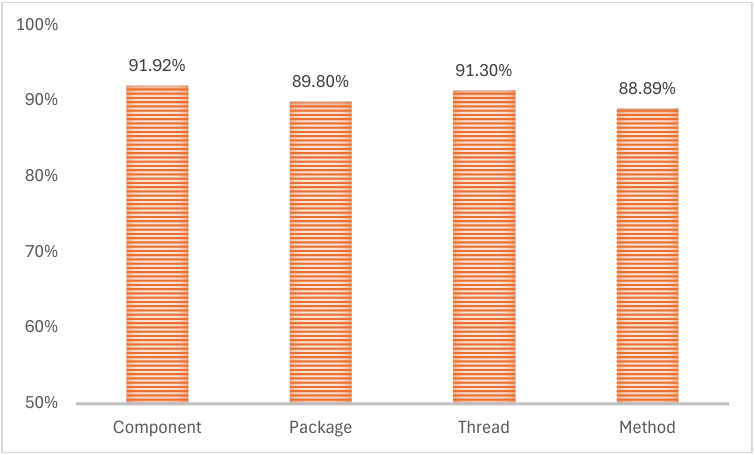}
    \caption{Accuracy of Hierarchy}\label{fig:hierarchy-accuracy}
\end{wrapfigure}

As shown in Table~\ref{table:ablation}, removing any component leads to a clear degradation in localization performance, confirming that all components contribute meaningfully to the effectiveness of ConFL.
Among the ablated variants, \textbf{w/o Hierarchy} suffers the largest performance drop, with Top-1 accuracy decreasing from 132 to 92.
This result highlights the critical role of hierarchical retrieval in progressively constraining the search space and guiding the LLM toward relevant concurrency contexts.
Without hierarchy, the model is more likely to explore irrelevant code regions and follow incorrect reasoning paths.

The \textbf{w/o CKB} variant also shows a substantial decline (Top-1 = 104), indicating that the concurrency knowledge base provides essential structural and behavioral signals beyond method signatures.
Without explicit knowledge of shared variables, locks, and thread-entry properties, the LLM’s ability to reason about concurrency interactions is significantly weakened.

In contrast, the \textbf{w/o DSL} variant exhibits a smaller but still noticeable performance drop (Top-1 = 117).
This suggests that while raw concurrency information is useful, the structured interaction-level DSL plays an important role in organizing this information into a form that is easier for the LLM to reason over.
By explicitly encoding thread entries, shared resources, and access patterns, the DSL improves ranking accuracy and overall localization quality.

Overall, the full ConFL framework consistently achieves the best Top-k results, demonstrating that hierarchical retrieval, the concurrency knowledge base, and the interaction-level DSL are complementary and jointly necessary for effective and explainable concurrent fault localization.



\paragraph{Accuracy of Hierarchy Retrieval.}
Figure~\ref{fig:hierarchy-accuracy} reports the accuracy of hierarchical retrieval at each stage of our framework.
Component-level accuracy measures whether the correct component is identified from the bug report, followed by package-, thread-entry-, and method-level accuracy measured conditionally at each subsequent stage.

\begin{wraptable}{r}{0.3\textwidth} \footnotesize
\vspace{-10pt}
\centering 
\caption{BoostN with CKB} \label{tab:boost}
\begin{tabular}{lcc}
\toprule
Technique & MRR & MAP \\
\midrule
BoostN &  0.103  & 0.099 \\
BoostN (CKB) &  0.167   &  0.165 \\
\bottomrule
\end{tabular}
\vspace{-10pt}
\end{wraptable}

As shown in the figure, ConFL achieves consistently high accuracy across all hierarchy levels, exceeding 88\% at every stage.
In particular, component-level and thread-entry retrieval exceed 91\% accuracy, indicating that both structural localization and concurrency-entry identification are highly reliable.
Accuracy decreases slightly at deeper levels, such as package and method selection, which is expected as the search space becomes more fine-grained.

This strong performance reflects the high-cohesion and low-coupling properties of software systems.
Components and packages typically encapsulate related functionality and concurrency responsibilities, enabling hierarchy-guided retrieval to progressively narrow the search space while preserving correct localization decisions.
Overall, these results demonstrate that hierarchical retrieval effectively leverages program modularity to support reliable interaction-level reasoning and LLM-based fault localization.

We emphasize that the gains of ConFL do not come solely from search-space reduction. Restricting retrieval to the CKB improves IR-based localization: as shown in Table~\ref{tab:boost}, BoostN(CKB) improves over BoostN from 0.103 to 0.167 MRR and from 0.099 to 0.165 MAP. However, the performance remains substantially lower than ConFL, indicating that filtering irrelevant methods alone is insufficient for concurrent fault localization. Similarly, the ablation study (Table~\ref{table:ablation}) shows that removing hierarchy-guided reasoning reduces Top-1 from 132 to 92 even under the same CKB setting. These results suggest that ConFL's gains mainly come from structured concurrency interaction modeling and hierarchy-guided reasoning rather than search-space reduction alone.

\vspace{-0.5em}
\vspace*{3pt}
\begin{tcolorbox}
\stepcounter{findingsCounter}
{\bf RQ\thefindingsCounter{} summary:} {
The ablation study confirms that hierarchical retrieval, the concurrency knowledge base, and the interaction-level DSL are all essential components of ConFL.
Hierarchical retrieval has the largest impact, as it effectively constrains the search space and guides the LLM toward relevant concurrency contexts.
This is further supported by the hierarchy-accuracy analysis, which shows consistently high accuracy across all levels, reflecting the high-cohesion and low-coupling structure of real-world software systems.
}
\end{tcolorbox}

%% file: tables/results/ablation.tex
\vspace{-4pt}
\begin{table}[H]\footnotesize
\caption{Ablation study} \label{table:ablation}
\vspace{-8pt}
\begin{tabular}{c|ccc|cc}
\hline \toprule
Variant & Top-1 & Top-3 & Top-5  & MRR & MAP \\ \hline
w/o Hierarchy  &  92     & 140      &  158 & 0.356   & 0.345     \\ 
w/o CKB    & 104      & 157      & 178     & 0.405   & 0.391   \\ 
w/o DSL    & 117      & 170      & 193     & 0.449   & 0.432     \\ \hline
ConFL     & 132      & 186      & 216    & \textbf{0.503}    & \textbf{0.486}     \\ \bottomrule
\end{tabular}
\end{table}
\vspace{-0.5em}

%% file: RQs/RQ4.tex
\subsubsection{RQ4: How well does our approach generalize to unseen data and different LLMs?}

To assess the generalization ability of our framework, we evaluate two complementary dimensions:Performance on unseen (post cut-off) bugs and Performance across different LLM backbones. These experiments test whether our system overfits to training-era bug reports, and whether the gains from our concurrency-aware design persist across model architectures.

\input{tables/results/results_post}

\noindent
\textbf{Generalization to Unseen Bugs.}
We evaluate our method on a post–cutoff dataset consisting of real-world concurrent bugs introduced after the LLM training cutoff, ensuring that none of the underlying models have prior exposure to these cases. As shown in Table~\ref{tab:postcutoff}, our framework maintains strong performance, achieving MRR = 0.615 and MAP = 0.598, substantially outperforming the LLM-based baseline FlexFL (MRR = 0.490, MAP = 0.477) as well as all IR-based methods, which fail to generalize in this setting. Notably, ConFL achieves 20 Top-1 hits out of 31, compared to FlexFL’s 14, demonstrating that our hierarchical retrieval and concurrency-aware reasoning remain effective even when applied to previously unseen systems and bug patterns. These results confirm that our approach generalizes well beyond the training-era distribution and remains robust under realistic post–cutoff conditions.

\begin{wrapfigure}{r}{0.35\textwidth}
    \centering
    \includegraphics[width=0.35\textwidth]{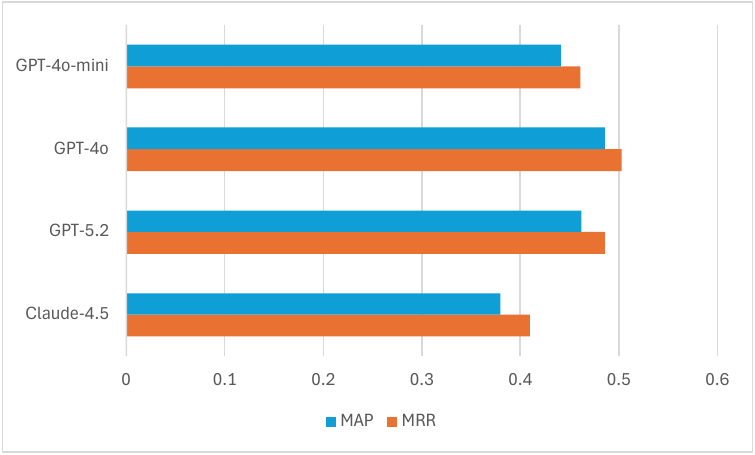}
    \caption{Performance across LLMs} \label{fig:llm-generalization}
    \vspace{-8pt}
\end{wrapfigure}

\noindent
\textbf{Generalization across different LLMs.}
We further evaluate the robustness of our framework using multiple LLM backbones, including GPT-4o, GPT-4o-mini, GPT-5.2 from OpenAI, and Claude~4.5 from Anthropic. Figure~\ref{fig:llm-generalization} shows that ConFL achieves its best performance with GPT-4o, while GPT-4o-mini delivers comparable results with only modest degradation. GPT-5.2 and Claude~4.5 exhibit slightly lower performance in some cases. Across all evaluated models, the MRR ranges from 0.41 to 0.503, corresponding to an absolute variation of 0.093, indicating relatively stable behavior across different LLM backbones.

Interestingly, newer or larger models do not consistently improve localization performance. We hypothesize that this behavior is related to differences in reasoning style and adherence to the hierarchy-guided retrieval process. Models such as GPT-4o and GPT-4o-mini appear to follow the structured interaction-level context more consistently, leading to more stable localization results.

Overall, these results suggest that ConFL benefits primarily from structured concurrency interaction modeling and hierarchy-guided reasoning rather than relying on a specific LLM architecture or unconstrained reasoning capability. The consistent improvements across different LLM backbones further demonstrate that the effectiveness of our framework stems from its structured reasoning process, making it robust and broadly applicable to a variety of foundation models.

\vspace{-0.5em}
\vspace*{3pt}
\begin{tcolorbox}
\stepcounter{findingsCounter}
{\bf RQ\thefindingsCounter{} summary:} {
The results show that ConFL generalizes well to both unseen concurrent bugs and different LLM backbones.
It consistently outperforms state-of-the-art baselines on post–cutoff bugs, indicating that its effectiveness does not rely on memorization.
Across LLMs, ConFL remains robust, with GPT-4o and GPT-4o-mini achieving the most stable performance.
Overall, these findings suggest that ConFL’s gains mainly come from hierarchy-guided retrieval and concurrency-aware interaction modeling, rather than from any specific LLM architecture.
}
\end{tcolorbox}

%% file: tables/results/results_post.tex
\vspace{-0.5em}
\begin{table}[H]\footnotesize
\caption{Performance on Dataset$_{Post}$} \label{tab:postcutoff}
\vspace{-8pt}
\begin{tabular}{c|ccc|cc}
\toprule
Technique & Top-1 & Top-3 & Top-5 & MRR & MAP \\ \hline
Flexfl    & 14      &   18    & 18       & 0.490   & 0.477    \\ \hline
ConFL     & 20      &  25     & 26      & \textbf{0.615}    & \textbf{0.598}    \\ \bottomrule
\end{tabular}
\end{table}
\vspace{-0.5em}

%% file: discussion.tex
\section{Discussion}
\subsection{\Name{}’s Failed Case Study}

Although \Name{} achieves strong overall performance, we observe several representative failure modes that reflect inherent challenges in concurrent fault localization and point to directions for future improvement.

\noindent
\textit{Hierarchy Retrieval Failure.}
\Name{} relies on hierarchical retrieval (component $\rightarrow$ package $\rightarrow$ entry point).
In some cases, incorrect components or packages are selected due to sparse documentation, misleading summaries, or missing concurrency-specific semantics (e.g., thread ownership or resource roles).
Once an early decision is wrong, subsequent retrieval is constrained to an incorrect scope.
Future work could mitigate this issue by enriching component-level summaries with explicit concurrency information and enabling limited backtracking or re-ranking across hierarchy levels.

\noindent
\textit{Third-Party Library Effects.}
Some failures arise when key concurrency behavior is encapsulated in third-party libraries such as Netty or Jetty.
Because \Name{} analyzes only project source code, it cannot observe implicit thread creation, callbacks, or scheduling logic implemented inside external libraries.
This may cause important execution paths to be missed.
Potential extensions include incorporating lightweight library summaries, LLM-generated abstractions, or bytecode-level signals.

\noindent
\textit{Noise in the Concurrent Knowledge Base (CKB).}
The CKB is built using over-approximate static analysis, favoring recall over precision.
As a result, it may introduce noise such as spurious call-graph edges, excessive shared-variable candidates, or imprecise aliasing, especially in polymorphic code.
This noise can reduce ranking stability and obscure the true fault.
We observe that manually refining the CKB (e.g., improving alias precision) leads to noticeable accuracy gains, suggesting that more precise static context would further strengthen \Name{}’s reasoning.

\subsection{Runtime and Token Cost}
CKB construction relies on lightweight static analysis based on JavaParser and Datalog rules, without expensive pointer or alias analysis. As shown in Table~\ref{tab:cost}, the average project in our benchmark contains 1,808 files and 14,252 methods, while the average CKB construction time is only 24.6 seconds. This indicates that the preprocessing overhead of ConFL remains low even for large-scale real-world systems.

\vspace{-0.5em}
\begin{table}[H] 
\centering
\caption{CKB Construction and LLM Cost}
\vspace{-8pt}
\label{tab:cost}
\scriptsize
\setlength{\tabcolsep}{5pt}

\begin{tabular}{cccccc}
\toprule
\textbf{Avg. Files} &
\textbf{Avg. Methods} &
\textbf{Avg. Build Time} &
\textbf{Avg. DSL Tokens} &
\textbf{Avg. LLM Calls} &
\textbf{Avg. Tokens/Bug} \\
\midrule
1808.1
 & 14251.6
 &  24.6(s)
 & 163.7
 & 8.94
 & 4261.23 \\
\bottomrule
\end{tabular}
\end{table}
\vspace{-0.5em}

The token cost during fault localization is also bounded. Instead of exposing large code regions or long call chains to the LLM, ConFL summarizes concurrency-relevant behaviors using compact interaction DSLs. On average, each interaction DSL contains 163.7 tokens, and the hierarchical retrieval process requires only 8.94 LLM calls per bug. As a result, the average token usage is 4,261 tokens per bug, remaining comparable to existing LLM-based fault localization approaches while providing substantially richer concurrency-aware context for reasoning.

\subsection{Performance on Different Concurrent Bug Types}
\label{sec:discussion-bugtype}

Our approach focuses on localizing concurrency faults based on observed symptoms described in bug reports.
Large language models are particularly effective at reasoning from such symptom-level information.
Introducing explicit bug-type labels (e.g., race condition, deadlock, etc) would add an additional source of bias, as concurrency bug classification itself is a challenging and often ambiguous task. 
Accordingly, our interaction-level DSL does not encode explicit bug-type categories.
Instead, it uniformly captures shared resources, execution contexts, and access actions that are directly relevant to concurrency behavior.
Our results suggest that fault localization performance is more strongly influenced by the quality of the bug report than by the nominal bug type.
When bug reports provide concrete concurrency-related information, localization accuracy remains consistently high regardless of the underlying fault category.
These observations motivate a holistic evaluation without enforcing potentially ambiguous concurrency bug classifications.

\subsection{Comparison with Other FL Techniques}
\label{sec:discussion-comparison}

We do not directly compare our approach with all existing fault localization techniques, as our problem setting differs fundamentally in terms of input assumptions and applicability.

Our approach operates using \emph{bug reports as the only input}. Under this constraint, applicable baselines are limited.  Traditional IR-based fault localization techniques fall into this category, as well as recent methods such as FlexFL. In contrast, many learning-based approaches require large, project-specific training datasets, often consisting of hundreds or thousands of historical bug reports.
In our setting, each project contains only tens of concurrency-related bug reports, making such approaches impractical.

Similarly, several recent LLM-based fault localization techniques (e.g., AutoFL and AgentFL) rely on dynamic execution information, such as test cases, coverage, or execution traces.
These methods are designed for settings where runnable test suites are available, which is not assumed in our work.
Our goal is to localize concurrency faults in scenarios where only bug reports and source code are accessible.

Other approaches, such as graph-based LLM reasoning methods, are often developed for Python projects and rely on constructing program graphs for reasoning.
Directly extending these techniques to large-scale Java systems is challenging due to the size and complexity of Java program graphs, including extensive virtual dispatch, framework-level abstractions, and dense dependency structures.
Constructing and reasoning over such graphs would significantly increase analysis overhead and is orthogonal to the lightweight, interaction-level abstraction adopted in our approach.

Taken together, these considerations motivate our comparison with IR-based and bug-report-driven baselines, which share the same input assumptions.
Our approach is complementary to execution-based and learning-based techniques, and is designed to address a distinct and practically relevant fault localization setting.

\section{Threats to Validity}
\label{sec:threats}

We discuss potential threats to the validity of our study from three perspectives.

\textbf{Internal Validity.}
Internal validity threats mainly stem from the use of large language models in the localization process.
Although LLMs may exhibit nondeterministic behavior, our approach mitigates this issue through hierarchical retrieval, constrained context, and deterministic decoding settings.
Nevertheless, different LLM implementations or inference configurations may still lead to variations in results.
To reduce such effects, we apply identical prompts, workflows, and decoding parameters across all evaluated models.

\textbf{External Validity.}
Our experiments are conducted on real-world Java projects with documented concurrency bugs. Although the entity-extraction layer is tailored to Java concurrency primitives, the benchmark includes diverse concurrency styles, ranging from lock-heavy systems (e.g., Redisson) to executor- and async-heavy systems (e.g., Pulsar). While adapting the extraction layer to other languages would require additional engineering effort, the hierarchy-guided retrieval and interaction-level reasoning components are largely language-agnostic.

\textbf{Construct Validity.}
Our interaction DSL captures shared-resource accesses and cross-thread interactions but does not model fine-grained synchronization semantics such as happens-before relations or memory-model effects. Consequently, the DSL is most effective for shared-access concurrency bugs such as races and atomicity violations, and less suitable for ordering-related bugs. We leave detailed subtype-specific evaluation to future work.

\section{Conclusion}

In this paper, we proposed \Name{}, an explainable fault localization framework for concurrent programs using bug reports alone.
By combining hierarchical retrieval with a concurrency-aware knowledge base and interaction-level DSL, \Name{} enables LLMs to reason over thread entries and shared-resource interactions in a structured and focused manner.
Extensive experiments on real-world concurrent bugs show that \Name{} consistently outperforms state-of-the-art IR-based and LLM-based approaches across different report qualities, unseen bugs, and LLM backbones.
These results indicate that effective fault localization depends more on structured concurrency modeling and guided retrieval than on raw LLM capacity.
Future work includes improving the precision of static concurrency knowledge, better handling concurrency behavior in third-party libraries, and integrating additional signals when available to further enhance robustness and explainability.